\documentclass{ws-ijmpd}

\begin{document}
\setcounter{page}{107} \markboth{H. Mosquera Cuesta, K.
Fiuza}{Neutrino Oscillations at Supernova Core Bounce Generate the
Strongest Gravitational-...}

\catchline{}{}{}{}{}

\title{Neutrino Oscillations at Supernova Core Bounce Generate the Strongest Gravitational-Wave Bursts  }

\author{\underline{Herman J. Mosquera Cuesta }}
\address{ Centro Brasileiro de Pesquisas F\'{\i}sicas \\ CEP 22290-180 Rio de Janeiro, RJ, Brazil
\\ hermanjc@cbpf.br}



\author{Karen Fiuza}
\address{Instituto de F\'{\i}sica, Universidade Federal do Rio Grande do Sul \\ CEP 91501-970 Porto
Alegre,RS, Brazil \\ kfiuza@if.ufrgs.br}

\maketitle

\begin{history}
\received{18 December 2003} \revised{23 January 2004}
\end{history}

\begin{abstract}
During the core bounce of a supernova collapse resonant
active-to-active  ($\nu_a \rightarrow  \nu_a$),  as well  as
active-to-sterile  ($\nu_a   \rightarrow   \nu_s$)   neutrino
($\nu$) oscillations can take place.  Besides, over this phase
weak magnetism increases  antineutrino ($\bar{\nu}$) mean free
paths, and thus its luminosity.  Because the oscillation feeds
mass-energy into the target $\nu$ species, the large mass-squared
difference between  species ($\nu_a \rightarrow \nu_s$) implies a
huge amount of power to be given off as  gravitational waves
($L_{\textrm{GWs}} \sim 10^{49}$~erg s$^{-1}$), due to anisotropic
but coherent $\nu$ flow over the oscillation length.  This
anisotropy in the $\nu$-flux is driven by both the {\it universal
spin-rotation} and the spin-magnetic coupling. The new spacetime
strain estimated this way is still several orders of magnitude
larger than those from $\nu$ diffusion (convection and cooling) or
quadrupole moments of the neutron star matter. This new feature
turns these bursts the more promising supernova gravitational-wave
signal that may be detected by observatories as LIGO,  VIRGO,
etc., for distances far out to the VIRGO cluster of galaxies.
\end{abstract}


\section{Generating GWs During $\nu$ Oscillations}

To start with, let us recall how the production of GWs during
$\nu$ oscillations proceeds by considering the case of
oscillations between active and sterile neutrinos in the supernova
core. The essential point here is that oscillations into sterile
neutrinos change dramatically the energy and momentum (linear and
angular) configuration of the proto-neutron star (PNS). In
particular, flavor conversions into sterile neutrinos drive a
large mass and energy loss from the PNS because once they are
produced they freely escape from the star. This means that
oscillations into sterile, in dense matter, take place over longer
oscillation lengths, compared to $\nu_a \rightarrow \nu_a$, and
the sterile encounter infinite mean free paths thereafter.
Physically, the potential, $V_s(x)$, for sterile neutrinos in
dense matter is zero, and their probability of reconversion, still
inside the star, into active species is quite small. This outflow
translates into a noticeable modification of the PNS mass and
energy quadrupole

Since most sterile neutrinos escape along the directions defined
by the dipole field and angular momentum vectors (see
Fig.\ref{4-dupole}), then the $\nu$ outflow is at least
quadrupolar in nature. This produces a super strong
gravitational-wave burst once the flavor conversions take place,
the energy of which stems from the energy and momentum of the
total number of neutrinos participating in the oscillation
process\footnote{The attentive reader must regard that neutrinos
carry away almost all of the binding energy of the just-born
neutron star, i.e., $\Delta E_\nu \sim 3\times 10^{53}$~erg.}.

The remaining configuration of the star must also reflect this loss,
i.e., the active $\nu$s were coupled to the neutron matter.
Hence its own matter and energy distribution becomes also quadrupolar.
Because this quadrupole configuration (the matter and  energy still
trapped inside the just-born neutron star) keeps changing over the
timescale for which most of the oscillations take place, then GWs
must also be emitted from the star over that transient. At the end,
the probability of conversion and the $\nu$-flux anisotropy parameter
($\alpha$, see below) determine both how much energy partakes in the
process and the degree of asymmetry during the emission.

The case for oscillations among active species is a bit different,
the key feature being that mass and energy is coherently relocated
(see Ref.\cite{mosquera-fiuza04}) from one region to another
inside the PNS, especially because of the {\it weak magnetism} of
antineutrinos that allows them to have larger mean free
paths\cite{horowitz02}. In addition, oscillations of electron
neutrinos into muon or tauon neutrinos leave these last species
outside their own neutrino spheres, and hence they are in
principle free to stream away.  This must also generate GWs during
that sort of flavor conversions, although their specific strength
(strain) must be a bit lower compared to conversions into sterile
neutrinos wherein almost all the $\nu$ species may participate,
and the large $\Delta m^2$ in the process. These neutral-current
interacting $\nu$ species must be the very first constituents of
the $\nu$ burst from any supernova since most $\nu_e$s are
essentially trapped.

Since the sterile neutrinos escape the core over a timescale of a
few milliseconds (ms), the number of neutrinos escaping and their
angular distribution is sensitive to the instantaneous
distribution of neutrino production sites. Since thermalization
cannot occur over such short times ($\Delta T^{\rm  thermal} \sim
0.5$~s), and since the neutrino production rate is sensitive to
the local temperature at the production site, the inhomogeneities
during the collapse phase get reflected in the inhomogeneities in
the escaping neutrino fluxes and their distributions. Because of
both the $\nu$ spin-$\vec{B}$ and $\nu$ spin-$\vec{J}$ coupling
the asymmetries in these distributions can give rise to quadrupole
moments, which must generate gravitational
waves\cite{mosquera00,mosquera02}, and dipole moments which can
explain the origin of pulsar kicks\cite{fuller03,loveridge}.

Fixed by the {\it probability of oscillation}, $P_{\nu_a \rightarrow
\nu_a}$, the {\it fraction} of neutrinos that can escape in the first
few milliseconds is, however, {\it small}. Firstly, the neutrinos have
to be produced roughly within one mean free path from their resonance
surface. Secondly, since in the case of ${\nu_a \rightarrow \nu_s}$
oscillations the $\nu_s$ is the heaviest neutrino species, the sign of
the effective potential $A(x)$ and the resonance condition indicates
that only ${\nu}_e$s and the antineutrinos $\bar{\nu}_\mu$ and
$\bar{\nu}_\tau$ can undergo resonant conversions.

\section{Anisotropic Neutrino Outflow: origin and computation}

To provide a physical foundation for the procedure introduced here to
determine the neutrino {\it asymmetry parameter}, $\alpha$, which
measures how large is the deviation of the $\nu$-flux from a spherical
one, we recall next two fundamental effects that run into action once a
PNS is forming after the SN collapse. We stress in passing that other
physical process such as convection, thermalization, etc., are not
relevant over the timescale under consideration: $\Delta T_{\rm osc}
\sim 3-10$~ms after the SN core bounce, and therefore do not modify in
a sensitive manner the picture described below. Indeed, if $\nu$
thermalization, for instance, already took place, then the oscillations
are severely precluded since oscillations benefit of the existence of
energy, matter density and entropy gradients inside the PNS
\cite{grimus03,akhmedov99}, which are ``erased'' once thermalization
onsets.

As discussed by Mosquera Cuesta and Fiuza\cite{mosquera-fiuza04} (see
also the detailed discussion by Loveridge\cite{loveridge} about the
$\nu$ distribution function inside the PNS, an asymmetric, off-centered
rotating $\nu$ beam) both the $\nu$-spin coupling to rotation and the
$\nu$-spin coupling to the magnetic ($\vec{B}$) field drives an effective momentum,
and thus energy flux, asymmetry along the
angular momentum ($\vec{J}$) and $\vec{B}$-field directions. The
combined action on the escaping $\nu$s of a rotating background
spacetime plus magnetic field makes their $\nu$-sphere a decidedly
distorted surface, with at least a quadrupolar distribution. This is
the source of the strong GWs burst in this mechanism when the
oscillations ensue.


Based on the concomitant action of both the effects: $\nu$-spin
coupling to both the magnetic field and rotation, described
previously, one can determine the $\nu$ flow anisotropy in a
novel, self-consistent fashion by defining $\alpha$ as the ratio
of the total volume filled by the distorted $\nu$-spheres to that
of the proto-neutron star ({\rm PNS}). The $\nu_e$-sphere radius
of a non-magnetic non-rotating star is obtained from the
condition: $$\tau_{\nu_e}(R_{\nu_e})       =
\int_{R_{\nu_e}}^\infty {\cal{K}}_{\nu_e} ~ \rho(r) ~ dr =
\frac{2}{3} \;,$$ where $\tau_{\nu_e}$ is the optical depth, and
${\cal{K}}_{\nu_e}$ the scattering opacity for electron neutrinos,
and $\rho$ the matter density.

One can take hereafter\cite{burrows95} $R^{\nu_e}_{\rm PNS} \equiv
R_{\nu} \sim 35$~km, which is of the order of magnitude of the
oscillation length $\lambda_{\nu}$ of a typical supernova $\nu$;
which is constituent as well of the atmospheric $\nu$s, for which
${\Delta m^2} \sim 10^{-3}  {\rm eV}^2$ has been estimated by
Superkamiokande $\nu$ detector\cite{fukuda98} $$ \lambda_{\nu}
\sim 31~{\rm  km}~ \left[\frac{ E_{\nu_e}}{10 ~{\rm MeV} }\right]
\left(\frac{ 10^{-3} ~{\rm eV}^2 }{\Delta m^2}\right)\;.$$
Therefore, resonant conversions between active species may take
place at the position $r$ from the center defined by $$r  =
R_{\nu_e} + \delta_0 \cos \phi\; , \label{nu-sphere-general}$$
with $\phi$ the angle between the $\nu$-spin and $\vec{B}$, i.e.,
$$\cos \phi = \frac{(\vec{k} \cdot \vec{B})}{\vec{k}} \, \, ; \,
\,  \delta_0 = \frac{e \mu_e B}{2 \pi^2 (dN_e/dr)} \sim 1-10~\rm
km\;,$$ for $B \sim 10^{14-15}$ ~G, respectively. Here $e$, $N_e =
Y_e N_n$ and $\mu_e$ represent the electron charge, density  and
chemical potential, respectively.

\begin{figure}
\label{4-dupole}
{ \centerline{ \includegraphics[width=3.5truein,angle=0]{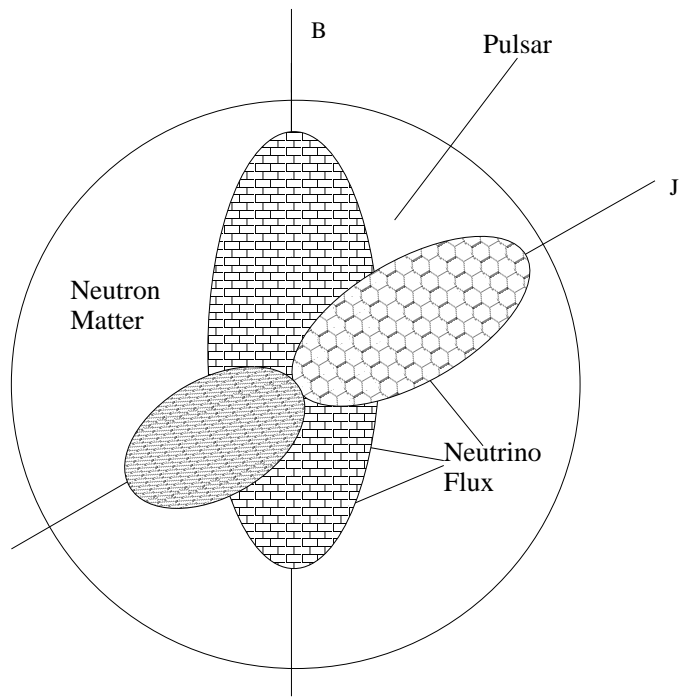} } }
\caption{A schematic representation of the $\nu$-flux
distribution (hatched regions) in a nascent pulsar. The at least
quadrupolar distribution is evident and stems from the neutrino
spin-magnetic and spin-rotation couplings.}
\end{figure}

This defines in Fig.\ref{4-dupole} an ellipsoidal figure of
equilibrium with semi axes $a  = R_\nu  + \delta_0; \hskip  0.5
truecm  {\rm and } \hskip 0.5truecm b = R_\nu$, and volume (after
rotating around $\vec{B}$): $$V_{ellips.} = \frac{4}{3}\pi R^2_\nu
(R_\nu + \delta_0)\;. \label{ellipsoidal-volume}$$

Meanwhile, the $\nu$-spin coupling to rotation generates an
asymmetric lemniscate-like plane curve (see Fig.\ref{4-dupole}) $
r = R_\nu~(1 \pm L~a~\cos~\theta)$\cite{mosquera-fiuza04}, which
upon a $2 \pi$ rotation around the star angular momentum axis
generates a volume: $$ V_{lemnisc.} = \frac{1}{4}(2 R_\nu)^2
\times 2 \pi \bar{y} + \frac{1}{4}(R_\nu)^2 \times 2 \pi \bar{y}\;
,$$ where the quantity $\bar{y}$ (and  $\bar{x}$) is defined as
the location of the centroid of one of the lobes of that plane
figure with respect to its coordinate center (x,y). After a long,
but straightforward, calculation one obtains $ \bar{y} =
\frac{\sqrt{\pi}}{\Gamma^2(1/4) }~R_\nu\; , \hskip 0.5 truecm
\bar{x} = \frac{4 \sqrt{\pi}}{\Gamma^2(1/4) }~R_\nu \; .$ Thence,
for a PNS\cite{burrows95} with parameters cited in this article
one obtains\cite{burrows96}

\begin{equation}
 \alpha_{\rm min} = \frac{ V_{ellips.} + V_{lemnisc.} }{ V_{\rm
PNS} } \sim 0.11-0.01\; . \label{param-alpha}
\end{equation}

As such, this is essentially a new result of this paper. The attentive
reader must notice in passing that the definition in
Eq.(\ref{param-alpha}) does take into account all of the physics of the
neutrino oscillations: luminosity, density gradients and angular
propagation, since Fig.\ref{4-dupole} does gather the relevant
information regarding the spatial configuration of the $\nu$ luminosity
in as much as is done in the standard definition of
$\alpha$\cite{burrows95,muller97}

\begin{equation}
\alpha(t)  \equiv \frac{1}{L_\nu (t)} \; \int_{4  \pi} d\Omega' \;
\Psi(\theta',\phi') \; \frac{L_\nu (\Omega',t)}{d\Omega'} \; .
\label{alpha-standard}
\end{equation}

Indeed, one can get the "feel" of the relationship between these two
definitions by noticing that the quantity $L_\nu (\Omega',t)$ in the
integrand of Eq.(\ref{alpha-standard}) can be expressed as $L_\nu
(\Omega',t) \equiv L_\nu(t) F(\Omega') $, where the function
$F(\Omega')$ contains now all the information regarding the angular
distribution of the neutrino emission. Hence $L_\nu(t) $ can be
factorized out of the integral and dropped  from
Eq.(\ref{alpha-standard}).  This  converts Eq.(\ref{alpha-standard}) in
a relationship among ({\it solid}) angular quantities, which clearly
can be reduced to a volumetric one, similar to the one introduced in
Eq.(\ref{param-alpha}), upon a transformation using the definition of
{\it solid angle} in the form of {\it Lambert's law}: $d\Omega =
\frac{dA \cos\theta}{R^2}$, and applied to the sphere representing the
PNS.  Here $\theta$, measured from a coordinate system centered at the
PNS\cite{muller97}, plays the role of the angle between the direction
towards the observer and the direction $\Omega'$ of the  radiation
emission in Eq.(24) of Ref.\cite{muller97}. Therefore, the novel result
here presented is physically consistent with the standard one for the
asymmetry parameter $\alpha$.

\section{Enlarged $\nu$ and GWs Luminosity From Oscillations in Dense
Matter}

Numerical simulations\cite{muller97} showed that, in general, the
fraction of the total binding energy emitted as GWs by pure $\nu$
convection\cite{mosquera00,burrows96,muller97} is: $E^\nu_{\rm GW}
\sim$ [10$^{-10}$-10$^{-13}$]~M$_\odot$c$^2$, for a $\nu$ total
luminosity: $L_{\nu} \sim 10^{53}$ erg~s$^{-1}$. In the production
of GWs via $\nu$ oscillations\cite{mosquera00} ($\nu_a
\leftrightarrow \nu_a$ or $\nu_a \leftrightarrow \nu_s$) there
exists two main reasons for expecting a major enhancement in the
GWs luminosity during the transition: a) the conversion itself,
which makes the overall luminosity ($L_{\nu_x}$) of a given
$\nu_x$ species to grow by a large factor: $\Delta L_{\nu_x}
\lesssim 10\% L^{\textrm{total}}_\nu $. The enhancement stems from
the mass-energy being given to, or drained from, the new  $\nu$
species into which oscillations take place. This augment gets
reflected in the species mass-squared difference $\Delta m^2$, and
their relative abundances: one species is number-depleted while
the other gets its number enhanced. But, even if the energy
increase, or give up, is small, b) the abrupt (resonant)
conversion over the transition time $\Delta  T_{\textrm{osc}}
\equiv \frac{  \lambda_{\textrm{osc}} } { \bar{V}_{\nu-Diff.} }
\sim [10^{-4}-10^{-3}]~\rm s , \label{oscillation-time}$ also
magnifies transiently $L_{\nu_x}$. Here $\lambda_{\textrm{osc}}$
defines the oscillation length, and $\bar{V}_{\nu-Diff.} \sim
10^9$~cms$^{-1}$ the convective $\nu$ diffusion
velocity\cite{muller97}.

If flavor transitions can indeed take place during supernovae
(SNe) core collapse and bounce, then they must leave some imprints
in the SNe neutrino spectrum. Main observational consequences of
neutrino conversions inside SNe include the partial or total
disappearance of the neutronization peak; the moment at which most
$\nu_e$s are produced, the interchange of the original spectrum
and the appearance of a hard $\nu_e$ spectrum, together with
distortions of the $\nu_e$ energy spectrum and alterations of the
$\bar{\nu}_e$ spectrum\cite{dighe}.

\subsection{Resonance, Adiabaticity and the Role of Weak Magnetism}

$\nu$ oscillations in vacuum produce no
GWs\cite{mosquera00,mosquera02}. In the case of active-to-active
$\nu$ oscillations (essentially the same physical argument holds
also for active-to-sterile $\nu$ oscillations), the main reason
for this negative result is that this class of conversions do not
increase in a significant figure the total number of particles
escaping from the proto-neutron star. In dense matter, the process
generates no GWs since the oscillations develop with the neutrinos
having very short mean free paths, so that they motion can be
envisioned as a standard diffusion process.

However, because of the active {\it antineutrino} species {\it
weak magnetism} their effective luminosity can be enlarged as much
as 15\% compared to the typical one they achieve when this effect
is not taken into consideration during their propagation in dense
matter\cite{horowitz02}. This result can be interpreted by stating
that the number of oscillating (and potentially escaping)
antineutrinos may be augmented by a large factor because now they
do encounter longer mean free paths. Below we take advantage of
this peculiar behavior of the $\nu$ outflow in supernovae in
computing the overall probability of transition between active
species.

\subsection{Active-To-Active $\nu$ Oscillations}

The neutrino oscillation pattern in vacuum can get noticeably
modified by the passage of neutrinos through matter because of the
effect of {\it coherent forward
scattering}\cite{wolfenstein79,smirnov85}. Therefore, interaction
with matter may help in allowing more $\nu$s to escape if resonant
conversions into active\cite{mosquera00} and/or sterile
$\nu$s\cite{mosquera00} occur inside the $\nu$-sphere of the
active $\nu$s. The transition probability\cite{grimus03} for
conversions among active species thus reads
\begin{equation} P_{\nu_a \rightarrow \nu_a}(|\vec{x}_1 - \vec{x}_0|) = 1 - \frac{1}{2}
\left[1 + \cos 2\theta_m (x_0) \cos 2\theta_m (x_1) \right],
\label{conversion-probability}
\end{equation}
where $x_0$ and $x_1$ correspond to the production and detection
sites, respectively, and $\theta_m $ the mixing angle in matter.

In order to produce an effect, neutrinos must be able to escape
the core without thermalizing with the stellar material. For
active neutrino species of energies $\approx 10$ MeV, this is not
possible as long as the matter density is $\gtrsim 10^{10}$
g~cm$^{-3}$. Since the production rate of neutrinos is a steeply
increasing function of matter density (production rate $\propto
\rho^n$, where $\rho$ is the matter density and $n>1$), the
overwhelming majority of the neutrinos of all species produced are
trapped.  This way, there seems to be no contribution to the GWs
amplitude for neutrino conversions taking place within the active
neutrino flavors. Nonetheless, if there were indeed {\it weak
magnetism effects}\cite{horowitz02}, one can rethink of
conversions within active species.  In the new result on {\it weak
magnetism for antineutrinos in core collapse supernovae}, the
antineutrinos ($\bar{\nu}_x$s) luminosity could be noticeably
increased because of their longer mean free
paths\cite{horowitz02}, and this means that the total energy flux
can be augmented in $\sim 15\%$ for $\nu$s of temperature $\sim
10$~MeV. One can verify that longer mean free paths allows for a
larger oscillation probability, and hence the contribution to the
generation of GWs during flavor conversions within active species
becomes non-negligible compared to the earlier case
\cite{mosquera00} where not weak magnetism effects were taken into
account.

For active-to-active $\nu$ flavor conversions, for instance:
$\nu_e \rightarrow \nu_\mu, \nu_\tau$; as implied by SNO results,
the resonance must take place at a distance $x_{\rm res}$ from the
PNS center and amid the active $\nu$-spheres, whenever the
resonance condition is satisfied $$4\Delta  m^2 \cos 2\theta_m =
2\sqrt{2} ~~G_F ~~N_e(x_{\rm res})~~ k_{\nu_e}\; ;$$ notice that
we neglected the magnetic field contribution. Here $k_{\nu_e} =
E_{\nu_e}/c$ is the $\nu_e$ momentum, and $N_e(x_{\rm res}) =
N_{e^-} - N_{e^+} \sim 10^{39-40}$~cm$^{-3}$ is the electron
number density. Thus the right-hand part of this equation reduces
to
\begin{equation}
2\sqrt{2} G_F E_{\nu_e}  N_e(x_{\rm res}) = 6.9 \times 10^{4}~{\rm
eV^2} \left[\frac{\rho}{10^{11} \rm g~cm^{-3}}\right]
\left(\frac{E_{\nu_e}} {\rm 10~MeV}\right) \; .
\end{equation}

For densities of order $\rho_{x = x_{res}} \sim
10^{10}$~g~cm$^{-3}$, i.e., recalling  that $\nu_e$s are produced
at the PNS outermost regions\cite{mosquera00} where the  electron
to baryon ratio is $Y_e \sim  0.1$, the resonance condition if
satisfied for $\Delta m^2 \sim 10^4$~eV$^2$, which implies a
neutrino mass of about $m_\nu \sim 10^2$~eV. Neutrino flavor
conversions in the resonance region can be strong if the
adiabaticity condition is fulfilled\cite{mosquera00}, i.e.,
whenever\cite{grimus03} $\frac{\Delta    m^2\sin^2 2\theta_m}{2
E_{\nu}\cos 2\theta_m} \left(\frac{1} {\rho} \frac{{\rm
 d}\rho}{{\rm  d}x}\right)^{-1}_{x = x_{\rm res}} \gg 1 \; ,
\label{adiabaticity-condition}$ where $x_{\rm res}$ is the
position of the resonance layer. Recalling that the typical scale
of density variations in the PNS core is $h_{N_e} \sim
(dN_e/dr)^{-1} \sim $~6~km, this adiabatic behavior could be
achieved as far as the density and magnetic field remain constant
over the oscillation length
\begin{equation}
\lambda_{\rm{osc}}    \equiv   \left(\frac{1}{\rho}
\frac{{\rm d}\rho}{{\rm d}x} \right)^{-1}_{x = x_{\rm res} }  \sim
\left(\frac{1}{2\pi} \frac{\Delta   m^2}{2  k_\nu} \sin(2\theta_m)\right)^{-1}
 \sim \frac{1~\rm cm}{\sin(2\theta_m)}\; ,
\label{oscillation-length}
\end{equation}
of  order $h_{N_e}$,  which  can  be satisfied  for  $\Delta m^2  \sim
10^4$~eV$^2$ as long as\cite{kusenko99} $\sin^2 2\theta_m > 10^{-8}.$
Although these $\nu$ oscillations could be adiabatic for a wide range
of mixing angles,  and thus a large number of $\nu$s could actually
oscillate, a $\nu_x$ mass such as this is incompatible with both viable
solutions to the Solar Neutrino Problem (SNP) and the most recent
cosmological constraints on the total mass of all stable neutrino
species that could have left their imprint in the Cosmic Microwave
Background Radiation (CMBR), as inferred from the observations
performed by the satellite WMAP: $m_\nu \sim 1$~eV. Therefore, we
dismiss this possibility since there appears to be no evidence for
neutrinos masses in this parameter range inside the PNS core.

On the other hand, if one takes into account the KamLAND
results\cite{kamland03}, which definitively demonstrated that $i$)
a large mixing angle (LMA) solution of the solar $\nu$ problem is
favored: $\sin^2 2\theta \sim 0.8$, $ii$) for a mass-squared
difference: $\Delta m^2 \sim 5.5 \times 10^{-5}$ ~eV$^2$ (we use
the approximate value $\Delta m^2 \sim 10^{-4}$~eV$^2$ for the
estimates below), one can see that resonant conversions with
$\Delta m^2 = m_{\nu_2}^2 - m_{\nu_1}^2 \sim 10^{-4}$~eV$^2$ would
take place in supernova regions where the density is about $\rho
\sim 10-30$~g~cm$^{-3}$, which corresponds to the outermost layers
of the exploding star.  Although a large number of $\nu$ species
can in effect participate of the transitions there, i.e, the
neutrino luminosity can be still a very large quantity, these
regions are of no interest for the gravitational-wave emission
from neutrino oscillations since the overall energy density at
that distance from the star center is very small. This does not
mean that no GWs are emitted from transitions there, it is to mean
that their strain is very small so as to be detectable.
Observations of $\nu$ oscillations\cite{dighe} in this parameter
range would provide useful information regarding the SNP, the
hierarchy of neutrino masses and the mixing $|U_{e 3}|^2$. Note in
passing that oscillations in this range would imply a mass for the
$\nu_2$ species $m_{\nu_2} \sim 10^{-2}$ ~eV ~$(\frac{\rho}{10~\rm
g~cm^{-3}})$, in the case when $m_{\nu_2} > m_{\nu_1}$. This
$m_{\nu_2}$ is compatible with current limits from
WMAP\cite{WMAP}.

Finally, let us consider oscillations in the parameter range estimated
from CMBR by WMAP observations. In this case, resonant $\nu$
transitions would take place in regions where the density is as high as
$\rho \sim 10^{7-8}$~g~cm $^{-3}$, that is, at the supernova mantle or
PNS upper layers. At these densities the oscillation length can be
still $\lambda_{\rm  osc} \sim 1-5$~km, and thus the conversions can be
considered as adiabatic. Thus the resonance condition can be satisfied
for $\Delta m^2 \sim 1$~eV$^2$ as long as $\sin^2 2\theta_m \lesssim
10^{-3}$.

Hence, by taking with this constraint, and recalling that  $i$) at
least 6 $\nu$-species can participate in the flavor transitions,
$ii$) most $\nu$s are emitted having $\vec{k}_\nu$ parallel to
$\vec{B}$, which implies a further reduction factor of 2, and
$iii$) most $\nu$s are emitted having $\vec{k}_\nu$ parallel to
$\vec{J}$ implying also an additional reduction factor of 2, one
can show from Eq.(\ref{conversion-probability}) that the
{\it{fraction}} of $\nu_a$ species that can eventually exchange
flavor during the first few milliseconds after core bounce turns
out to be\cite{mosquera00} $P_{\nu_a \rightarrow
\nu_a}(|\vec{x}-\vec{x}_{res}|) \lesssim 0.1~\%\;.
\label{prob-active-conv}$ Thus, the total energy involved in the
oscillation process we are considering could be estimated as:
$E_\nu^{\rm tot} \simeq P_{\nu_a \rightarrow \nu_a} (|\vec{x} -
\vec{x}_{res}|) \times N_\nu  ~k_\nu ~c ~(1 + 0.15)$, with $N_\nu$
the total number of neutrinos undergoing flavor conversions during
the timescale $\Delta T_{\rm osc}$.  Note in passing that KamLAND
$\bar{\nu}_e$ experiment suggests\cite{kamland03} $P_{\bar{\nu}_e
\rightarrow \bar{\nu}_{\mu,\tau} } \sim  40\% $, while LSND $P_{
\bar{\nu}_\mu \rightarrow \bar{\nu}_e } \sim 0.26\%$. This result
from LSND has not been so far ruled out by any terrestrial
experiment, and there is a large expectation that it could be
verified by MiniBoone at Fermilab.

In the case of active-to-sterile $\nu$ conversions, it was proved
in Refs.\cite{mosquera00,mosquera02,mosquera-fiuza04} (see details
therein) that a substantial fraction: $P_{\nu_a \rightarrow
\nu_s}$, of $\nu$s may get converted to sterile $\nu$s, and
escape the core of the star, if the sterile $\nu$ mass
($m_{\nu_4}$) is such that

\begin{equation}
 1~{\rm  eV}^2  \lesssim  \Delta  m^2_{{\nu}_a  \rightarrow {\nu}_4}
\lesssim 10^4~{\rm eV}^2 \; . \label{solution}
\end{equation}

For both classes of $\nu$ conversions the number of $\nu$s escaping and
their angular distribution is sensitive to the instantaneous
distribution of production sites. These inhomogeneities can give rise
to quadrupole moments that generates GWs\cite{mosquera00,muller97}, and
dipole moments that could drive the runaway pulsar
kicks\cite{fuller03,loveridge}. Noting that at least 6 $\nu$-species
can participate in both types of oscillations and that the interaction
with both the magnetic field and angular momentum of the PNS brings
with an overall reduction factor of 4 in the oscillation probability,
one  can show that  the {\it{fraction}} of $\nu$s that can actually
undergo flavor transitions in  the first few milliseconds
is\cite{mosquera00}

\begin{equation}
P_{\nu_a   \rightarrow    \nu_s}^{\nu_a    \rightarrow   \nu_a}
(|\vec{x}-\vec{x}_0|) \lesssim 1\% , \label{prob-conv-sterile}
\end{equation}

of the total $\nu$s number: M$_\odot \times  m_p^{-1} \sim 10^{57} \nu$,
which corresponds to a total energy exchanged during the transition

\begin{equation}
\Delta  E_{\nu_a \longrightarrow \nu_s}^{\nu_a  \rightarrow \nu_a}
\sim 3\times 10^{51}~\rm erg . \label{energy-total}
\end{equation}

Equations (\ref{energy-total},\ref{oscillation-time}) determine the total
luminosity of the neutrinos participating in the resonant flavor conversions.

\section{GWs Energetics From $\nu$ Luminosity and Detectability}

If $\nu$ oscillations do take place in the SN core, then the most
likely detectable GWs signal should be produced over the time
interval for which the conditions for flavor conversions to occur
are kept, i.e., $\sim(10^{-1}-10)$~ms. This timescale implies GWs
frequencies in the band: $f_{\rm{GWs}} \sim$ [10 - 0.1]~kHz,
centered at  1~kHz, because of the maximum $\nu$ production around
$1-3$~ms after core bounce\cite{mosquera00}. This frequency range
includes the optimal ban width for detection by ground-based
observatories. For a 1~ms conversion time span the $\nu$
luminosity reads

\begin{equation}
L_{\nu}  \equiv   \frac{\Delta  E_{\nu_a  \longrightarrow  \nu_s}
 }{\Delta  T_{\textrm{osc}}  }  \sim  \frac{3\times 10^{51}  \rm  erg}
 {1\times  10^{-3} \rm  s}  = 3  \times  10^{54} \frac{\rm  {erg}}{\rm
 s}\;. \label{nu-luminosity}
\end{equation}

To estimate the GWs burst amplitude produced by the non-spherical
outgoing front of oscillation-produced $\nu_s$ one can write (see
details in
Refs.\cite{mosquera-fiuza04,mosquera00,burrows96,muller97})

\begin{eqnarray}
 h_\nu  =  \frac{2G}{c^4  R}  \left(\Delta t  \times L_\nu  \times
\alpha\right) \simeq  |A| \left[\frac{55\rm kpc}{R}\right]
\left(\frac{\Delta  T   }{10^{-3}  \rm  s}\right)  \left[\frac{L_\nu}{
3\times     10^{54}\frac{\rm     erg}     {\rm    s}     }     \right]
\left(\frac{\alpha}{0.1}\right) \; , \nonumber \label{burrows96a}
\end{eqnarray}
where $|A| = 4 \times 10^{-23}~ {\rm Hz}^{-1/2}$ is the amplitude.
\footnote{One must realize at this point that the high  value of the
anisotropy parameter here used is consistently supported by the
discussion by Mosquera Cuesta and Fiuza\cite{mosquera-fiuza04}
regarding the neutrino coupling to rotation and magnetic
field quoted above.} A GWs signal this strong will likely be
detected by the first generation of GWs interferometers as LIGO, VIRGO,
etc. Its imprint in the GWs waveform may resemble a spike of high
amplitude and time width of $\sim$~ms followed  by a Christodoulou's
memory\cite{mosquera02}. From Eq.(\ref{burrows96a}) the GWs luminosity
turns out to be

\begin{equation}
L_{\textrm{GWs}} \sim 10^{48-50} ~~ \frac{\rm {erg}}{\rm s}
~~  \left[\frac{L_\nu}{3\times  10^{54} \frac{\rm {erg}}{\rm  s}}
\right]^2 \left(\frac{\alpha}{0.1-0.01}\right)^2\;,
\end{equation}
while the GWs energy radiated in the process yields $ E_{\textrm{GWs}}
\equiv   L_{\textrm{GWs}} \times \Delta T_{\textrm{osc}} \sim
10^{47-45}~{\rm erg}\;.$ This is about $10^{5-3} \times L^{\rm
Diff.}_\nu$ the luminosity from $\nu$ diffusion inside the PNS
\cite{mosquera02}.

\section{Conclusion}

One can see that if $\nu$ flavor conversions indeed take place
during SN core bounce inasmuch as they take place in our Sun and
Earth\cite{smirnov02}, then GWs should be released during the
transition. The GWs signal from the process is expected to
irradiate much more energy than current mechanisms figured out to
drive the NS dynamics at birth do. A luminosity this large
(Eq.(\ref{nu-luminosity})) would turn these bursts the strongest
GWs signal to be detected from any SN that may come to occur on
distances up to the VIRGO cluster, $R \sim [10-20]$ Mpc. It is
stressed that this signal will still be the stronger one from a
given SN, even in the worst case in which the probability of $\nu$
conversion is three orders of magnitude smaller than the estimated
in the present paper. In proviso, we argue that a GWs signal that
strong could  have been detected during SN1987a from the Tarantula
Nebula in the Milky Way's satellite galaxy Large Magellanic Cloud,
despite of the low sensitivity of the detectors at the epoch. In
such a case, the GWs burst must have been correlated in time with
the earliest arriving neutrino burst constituted of some active
species given off during the very early oscillation transient
where some $\nu_e$s went into $\nu_{\mu}$s, $\nu_\tau$s or
$\nu_s$s. Thenceforth, it could be of worth to reanalyze the data
collected far from that event taking careful follow up of their
arrival times, if appropriate timing was available at that moment.

\section{Acknowledgements}

{The authors are indebted to Prof. C\'esar Vasconcellos
for inviting us to participate in this important meeting, which make it possible to start with
this our collaboration. HJMC acknowledges the support from FAPERJ, Brazil, through the grant
E-26/151.684/2002. KF thanks CAPES (Brazil) for a graduate fellowship.}

\end{document}